# Deep Learning for Space Weather Prediction: Bridging the Gap between Heliophysics Data and Theory

John C. Dorelli[1], Chris Bard[1], Thomas Y. Chen[2], Daniel Da Silva[1,3], Luiz Fernando Guides dos Santos[1,4], Jack Ireland[1], Michael Kirk[1,5], Ryan McGranaghan[1,5], Ayris Narock[1,6], Teresa Nieves-Chinchilla[1], Marilia Samara[1], Menelaos Sarantos[1], Pete Schuck[1], Barbara Thompson[1]

1. NASA, Goddard Space Flight Center, Greenbelt, MD
2. Columbia University, New York, NY
3. Trident Vantage Systems, LLC, Arlington, VA
4. Catholic University of America, Washington, DC
5. Atmosphere and Space Technology Research Associates, Louisville, CO
6. ADNET Systems, Inc., Greenbelt, MD

***ABSTRACT*** *Traditionally, data analysis and theory have been viewed as separate disciplines, each feeding into fundamentally different types of models. Modern deep learning technology is beginning to unify these two disciplines and will produce a new class of predictively powerful space weather models that combine the physical insights gained by data and theory. We call on NASA to invest in the research and infrastructure necessary for the heliophysics' community to take advantage of these advances.*

***1. NASA's place in the rapidly evolving machine learning landscape.*** The last decade has been a period of explosive growth in the use of machine learning (ML), and deep neural networks (DNN) in particular, to solve problems of practical interest (e.g., see LeCun *et al.*, [*Nature*, 521, 2015] for a summary of how recent advances in deep learning have enabled it to solve many real world problems). New developments in training algorithms, programming environments and hardware have made sophisticated deep learning architectures accessible to the non-specialist. Any scientist can now construct a complex neural network model with fewer than a hundred lines of code and run it on a cluster of Graphics Processing Units (GPUs) using freely available software packages like PyTorch (https://www.pytorch.org) and Tensorflow (https://www.tensorflow.org). This crossover of deep learning technology into the general science community has resulted in a breadth of exploration and experimentation that would not have been possible even a few years ago. A few recent examples in the heliophysics community include the use of Convolutional Neural Networks (CNN) to infer the vector magnetic field in the Sun's photosphere from spectropolarimetric data [Liu *et al.*, *Ap.J,* 894, 2020], classify the internal structure of Interplanetary Coronal Mass Ejections (ICME's) [Dos Santos *et al.*, *Solar Physics,* submitted, 2020], and predict magnetospheric substorm onset [Maimaiti *et al.*, *Space Weather,* 17, 2019]. Many other examples from the ML-Helio 2019 Conference in September of 2019 [Camporeale, *J. Geophys. Res.,* 125, 2020] are collected at https://ml-helio.github.io.

NASA has taken notice and has invested in several efforts to use modern data science, ML and artificial intelligence (AI) methods to support its science and engineering efforts. The Center for Data Science and Technology (https://datascience.jpl.nasa.gov) coordinates such efforts at the Jet Propulsion Laboratory (JPL), including a Data Science Working Group (established in 2014) to explore use cases and support pilot projects. The Frontier Development Lab (https://frontierdevelopmentlab.org/), funded by NASA through a cooperative agreement with the SETI Institute, aims to build public-private partnerships (e.g., by pairing ML experts with scientists) to apply AI to problems of public interest. The Center for HelioAnalytics (CfHA) at NASA-GSFC has recently received NASA funding to initiate pilot projects that seek to enhance discovery in the Heliophysics domain.

It is clear that NASA recognizes the need to build connections with the ML and AI communities to support its mission. What is less clear is how NASA's role in this broad and rapidly changing landscape will evolve in the next decade and beyond. We argue that the enormous quantity of data that will be generated by

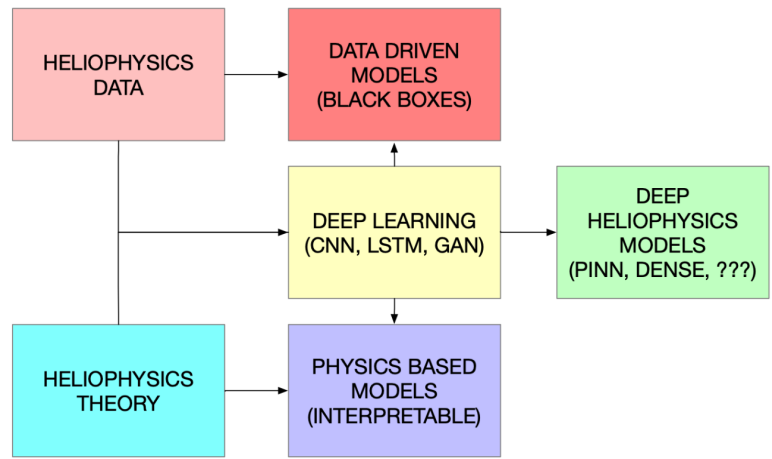

*Figure 1. Deep learning is changing the way we view the distinction between data driven and physics based models. Traditionally, data analysis and theory have been viewed as separate disciplines (salmon and turquoise boxes), each feeding into fundamentally different types of models (red and lavender boxes). Modern deep learning technology (yellow box) has begun to unify these two disciplines and will produce a new class of Deep Heliophysics Models (green box) that combine the physical insights gained by data and theory.*

NASA's heliophysics missions in the next several decades will make high performance deep learning (the training of deep neural networks with billions of parameters on large supercomputers) the dominant method for constructing models with data in the same way that high performance plasma simulation has become the dominant method for constructing physics based models. Further, we believe that high performance deep learning is about to transform the way we think about the development of space weather models, providing a bridge between "physics based" and "data driven" modeling that will enable significant progress in both areas (see Figure 1).

**2. Deep learning for space weather prediction.** Most of the current research on deep learning applications in heliophysics focuses, naturally, on space weather prediction. In this context, deep learning models are often viewed as complicated "black boxes," more akin to empirical models than to physics based models. This prevailing view often makes scientists hesitant to adopt these techniques. Experimentalists may ask, "What is gained by replacing well-established empirical models with more complex ones that require specialized knowledge, software and high performance computing to build and deploy?" Theorists and physics based model developers may ask, "What generalizable physical insights can be extracted from this black box that can compete with the knowledge we gain from our physics based models?" The answer to the first question will become clear as more applications show significant improvements over older methods (the Stokes inversion problem [Liu *et al.*, *Ap.J.,* 894, 2020] is a good example). The second question is often motivated by the assumption that neural networks are uninterpretable, providing predictive power with little physical insight. We argue that recent developments in the use of deep neural networks to simulate physical systems are blurring the traditional boundaries between "physics based" and "empirical" modeling.

The great challenge for physics based space weather models is the vast range of scales that must be addressed. Brute force kinetic simulations are out of the question and will remain so for the foreseeable future. Even fluid simulations are expensive, making ensemble forecasting impractical. Can deep learning help? Viewed as universal function approximators (e.g., Balázs [MSc Thesis, Eötvös Loránd University, 2001]), neural networks would seem to be ideal models to discretize partial differential equations (PDEs). Early attempts (e.g., Lagaris *et al.* [*IEEE Transactions on Neural Networks,* 9, 1998]) were successful on simple problems, achieving accuracy comparable to standard methods for simple systems with smooth solutions; but work stagnated for the next two decades due to the difficulties in training DNN with many hidden layers.

With the renaissance of deep learning in the mid-2000's came a resurgence of research on using DNN to solve PDEs. Will recent DNN advances enable the simulation of complex multiscale systems (like the Sun's atmosphere or Earth's magnetosphere) consisting of relatively smooth regions with embedded kinetic scale structures? Deep neural networks should excel at representing such systems. For example, the CNN architecture, being closely related to sparse coding [Oldhausen *et al.*, *Vision Res.,* 37, 1997; Papyan *et al.*, *J. Mach. Learn. Res.,* 18, 2017], is well suited to modeling the statistics of natural images even in the absence of millions of training examples [Ulyanov *et al, arXiv:1711.10925v4,* 2020]. In general, DNN can be viewed as efficient, multiscale basis function expansions, with each successive layer building more complex basis functions from the more primitive ones in the previous layer. Thus, one might expect that DNN can be used for the efficient solution of partial differential equations by collocation methods. Indeed, recent efforts to develop "Physics Guided Neural Networks" (PGNN) [Karpatne *et al.*, *IEEE Trans. on knowledge and data eng.,* 29, 2017] and "Physics Informed Neural Networks" (PINN) [Raissi *et al.*, *J. Comp. Phys.,* 378, 2019] are promising. The PINN approach is particularly interesting for two reasons: 1) it uses automatic differentiation (available through relatively simple interfaces in both PyTorch and Tensorflow) to incorporate PDE constraints *exactly* into the loss function at training samples, 2) it has been extended to architectures like CNN that are tailored to modeling multiscale systems.

An alternative to PINN is the "physics emulator" approach, in which one trains a DNN on a large number of physics based simulations and then replaces the physics based simulation with the DNN. The problem with this approach is obvious: large scale physics based simulations are expensive, and running a million or more to train a DNN is not practical (though the resulting simulation efficiency gains could motivate the archiving of simulation output for future use in training DNN). Recently, however, it has been shown how to use the inherent ability of CNN to model natural signals [Ulyanov *et al.*, *arXiv:1711.10925v4,* 2020] to dramatically reduce the number of physics based simulations needed to train a physics emulator. The approach [Kasim *et al.*, *arXiv:2001.08055v1,* 2020], known as Deep Emulator Network Search (DENSE), employs an algorithm known as Efficient Neural Architecture Search [Pham *et al.*, *arXiv:1802.03268v2,* 2018] to optimize the CNN architecture for a given problem domain. Dramatic reductions in the number of simulations required to train the emulator have been achieved with this approach.

While the PINN and DENSE approaches represent significant advances in the use of deep learning to model physical systems, we are clearly still at the start of the journey. How can more recent developments in deep learning – e.g., Long Short Term Memory (LSTM) architectures for sequence prediction problems (like natural language processing) or adversarial training methods that have proven so effective at randomly

*generating* complex data (like natural images or human faces) – be used to efficiently simulate physical systems? How do we effectively use data to constrain our physics based models? With data constraints added to the PINN loss function, for example, we can view the physics constraints as "regularizers" that prevent overfitting of a small number of training samples. More research is needed to determine the optimal way to combine data and physics constraints in DNN loss functions.

Ultimately, deep learning will produce powerful new methods for combining physics knowledge with data to build efficient, predictively powerful space weather models (Figure 1). To enable the heliophysics community to fully realize this potential, NASA should increase investment not just in the traditional top and bottom paths of Figure 1. In addition to continuing to encourage machine learning efforts through the TMS (Theory, Modeling and Simulations) and Science Center elements of the Heliophysics Grand Challenges program, NASA should invest in two critical areas: 1) more agile research programs that explore how the latest deep learning technology can be used to combine data and models in novel ways; 2) the computational infrastructure (including large GPU clusters, similar to those supported by the Department of Energy) to enable the training and deployment of large scale deep learning models. The next decade will be an exciting time as new advances in high performance deep learning become more widely available. We call on NASA to continue to encourage and invest in the research necessary for the heliophysics community to benefit from these advances.